\documentstyle[aps,epsfig,color%
]{revtex}

\newcommand{\beq}{\begin{equation}}
\newcommand{\eeq}{\end{equation}}
\newcommand{\beqa}{\begin{eqnarray}}
\newcommand{\eeqa}{\end{eqnarray}}
\newcommand{\id}{\mbox{{\small 1}\hspace{-0.37em}1}}
\newcommand{\df}{{\mathrm d}}
\newcommand{\tr}{{\mathrm tr}}
\newcommand{\intd}[2]{\int\!\!\frac{\df^{#1}{#2}}{(2\pi)^{#1}}}
\newcommand{\ket}[1]{|#1\rangle}
\newcommand{\bra}[1]{\langle #1|}
\newcommand{\sla}[1]{\slash\hspace{-0.6em}#1}

\begin{document}

\title{\Large\bf\sc Weak Decays Of Heavy Mesons In A Covariant Quark Model}

\author{Dirk Merten, Ralf Ricken, Matthias Koll, Bernard Metsch, Herbert Petry\\\vspace*{0.25cm}}

\address{Institut f\"ur Theoretische Kernphysik,\\ Nu{\ss}allee 14--16,\\D-53115 Bonn,\\Germany\\e-mail: merten@itkp.uni-bonn.de\\}

\date{\today}

\maketitle

\abstract
In this paper we investigate weak decays of heavy mesons in the
framework of a covariant quark model, which is based on the Bethe-Salpeter
 equation in instantaneous approximation. Apart from a
phenomenological confinement potential, a residual interaction induced
by instantons is adopted. Masses and many decay observables of
light mesons have already been described successfully in this model
\cite{Koll,Ralf}. An appropiate extension allows a unified
description of light and heavy systems.

Using a set of parameters which are fixed by the mass spectra, we
evaluate the form factors of semileptonic decays of charmed and bottom
mesons. In the heavy quark limit these can be reduced to the
Isgur-Wise function, which is calculated. Finally the form factors are
used to determine the non-leptonic decay rates of B mesons in the
factorization approximation.   
\endabstract

\setlength{\parskip}{1ex}

\section{Introduction}

In the last few years new and improved data on the spectra and the
decays of charmed and bottom mesons have become available. The
observations of the $D_1^ *$ and the radial excited ${D^{*\prime}}$ and $B^\prime$ have
 been published recently (see \cite{Ciulli} and references therein). The dominant semileptonic decays of $B$, $D$
and $D_s$ mesons have been measured with good precision  in the
meantime, and data for double Cabibbo-suppressed channels are also
available. Yet many new results will be provided by the B-factories {\sc BaBar}, {\sc Belle}, {\sc Hera-B} and {\sc LHC-B}
  within the next years.

For the theoretical description of these masses and decays various ans\"atze are pursued. Lattice gauge theory gives good
results for the transition form factors for high momentum transfer
$q^2$, while QCD sum rules are suitable for the low $q^2$ regime.
Heavy quark effective theory (HQET), which is based on a systematic
expansion in the inverse mass of the heavy quark, provides some model independent predictions, {\it i.e.}
approximate degenerated mass doublets according to different
orientation of the heavy quark spin,
and a connection between the semileptonic partial decay rate of the
pseudoscalar and pseudovector decay channels,
which has been experimentally confirmed. Unfortunately it
can not predict the masses and decay rates itself, and corrections due
to the finite quark mass, at least for the charm quark, are expected
to be  substantial. Hence for a consistent description of meson masses and decays over the full
kinematic region, constituent quark models, even if the connection
to the underlying theory is 
not quite clear, are still the most successful tool.

In previous papers we have developed a relativistic constituent quark
model for light mesons, which is based on the Bethe-Salpeter equation
in instantaneous approximation. Apart from a phenomenological confinement
potential we adopt a residual interaction induced by instantons. In
this model, a very good description of the 
light meson masses, from the ground state nonet up to highest angular momenta, has been
achieved. Also many decay observables have been calculated in reasonable
agreement with the experimental data (see \cite{Koll} for a recent update). Motivated by this success, the
model has been extended for heavy flavours \cite{Andre}. Thereto we do
not apply the One Gluon Exchange, which is known to give a good
description of heavy quarkonia and even of the whole meson spectrum
with a suitable inclusion of relativistic effects \cite{Godfrey}. Instead we
generalize the instanton induced interaction in a naive way\footnote{For a similar investigation of weak meson decays in the Bethe Salpeter framework  with the One Gluon Exchange see \cite{ZHMB}.}. This is
done in order to keep the model simple and unified: the
parametrization of confinement should be valid for all flavours, and
only one residual interaction is introduced. Also the question should
be raised, when and how the model for light mesons fails, if higher
quark and meson masses are involved. It turns out, that
heavy-light mesons and, to a certain extent, heavy quarkonia can be
described in that way.  The resulting
spectra together with a brief review of our model are discussed in
section II. Knowing the meson amplitudes we then calculate the
semileptonic decays without further parameters in section III. This is done in order
to test these amplitudes rather than to determine decay rates or CKM
matrix elements with high precision. Therefore we concentrate on the
dominant processes $B\to D^{(*)}\ell\bar\nu$ and on the CKM-favoured  
decay channels of $D$ and $D_s$ mesons. Finally in section IV we
investigate non-leptonic decays of B mesons, and section V contains our
summary.

\section{The Bethe-Salpeter Model}

Since the model has been described earlier in greater detail
\cite{BSE1,BSE2}, we will only briefly review the main features. 
The model is based on the Bethe-Salpeter equation for $q\bar q$ bound
states 
\beqa
{\chi}^P(p)=-i\;S_1^F({ \frac P2 +p})\intd{4}{p'}{ K}({ P;p,p}'){\chi}^P({ p}')S_2^F({ -\frac P2 +p'})
\eeqa
for the Bethe-Salpeter amplitude $\chi^P(p) :=
\bra{0}T\psi(p)\bar\psi(p)\ket{P}$. Here $\ket{P}$ denotes a bound state with mass $M$ and total momentum $P$, $P^2=M^2$.   
In our ansatz the full quark propagators $S_i^F$ are
approximated by free fermion propagators $ S^F_i({p}) \approx
i/\left(\sla{p}-m_i+i\varepsilon\right)$, where the constituent
quark masses $m_i$ are introduced, which are treated as free
parameters of the model. Furthermore, the irreducible interaction kernel $K$ is assumed to
be instantaneous in the rest frame of the meson,
${K}(P;p,p')|_{P=(M,\vec 0)} = {V}(\vec p, \vec p\,')$. These assumptions
lead to the (full) Salpeter equation 
\beq
{\Phi}(\vec p) =  \Lambda_1^-(\vec p)\gamma^0\left[\intd{3}{p'}\frac{{ V}(\vec p, \vec p\,'){\Phi}(\vec p\,')}{M+\omega_1+\omega_2}\right]\gamma^0\Lambda_2^+(-\vec p) - \Lambda_1^+(\vec p)\gamma^0 \left[\intd{3}{p'}\frac{{ V}(\vec p, \vec p\,'){\Phi}(\vec p\,')}{M-\omega_1-\omega_2}\right]\gamma^0\Lambda_2^-(-\vec p)
\eeq
for the Salpeter amplitude  ${ \Phi}(\vec p) := \int \frac{\mbox{\footnotesize d}p^0}{2\pi}\chi^P(p)\vert_{ P=(M,\vec 0)}$. Here $\omega_i := \sqrt{\vec
p\,^2+m_i^2}$ and $\Lambda^\pm$ denotes the projector on
positive and negative energy solutions of the Dirac equation. 

As ansatz for the interaction kernel we apply the sum of a
phenomenologically motivated confinement potential and a residual
interaction, which is induced by instantons. Confinement is
parametrized by a linearly rising potential in configuration space
with an adequate Dirac structure, symbolically written as 
$$V_C(r) = (a_c + b_c \cdot r)\;\mit\Gamma\otimes\mit\Gamma.$$
To estimate the influence of the Dirac structure, two possibilities are
taken into account, in the following referred to as model $\cal A$ and
model $\cal B$: a combination of scalar and time-like vector, which
has already been discussed in \cite{BSE1,BSE2} and which is
known to minimize spin-orbit splittings, and a chirally invariant
combination of scalar, pseudoscalar and vector type, previously
investigated by \cite{Gross} and \cite{Boehm}.

The additional instanton induced residual interaction  is based on the work of 't Hooft \cite{Hooft} and Shifman {\it et al.}\cite{Shifman}. The
effect of QCD instantons on the quarks leads to an effective two body
interaction, which can be expressed by the
following  kernel in momentum space 
\beq
\intd{3}{p'} V_{\mbox{\tiny III}}(\vec p,\vec p\,')\Phi(\vec p\,')  =  4\,{G}\, \intd{3}{p'}R_{\Lambda}(\vec p,\vec p\,')\cdot\left(\tr\left[\id \Phi(\vec p\,')\right] \id + \tr\left[\gamma^5 \Phi(\vec p\,')\right]\gamma^5\right).\label{III}
\eeq
The  strengths of the coupling are collected in  the flavour coupling matrix 
\beq {G}_{f_2f_3,f_1f_4} = \left\{\begin{array}{rcl}
                                              -g_{f_1f_2} & : & f_1=f_3 \neq f_2=f_4\\
					       g_{f_1f_3} & : & f_1=f_2 \neq f_3=f_4\\
						0         & : & \mathrm{otherwise},
					       \end{array} \right.\label{IIImat}
\eeq
with flavour indices $f_i$. The originally point like interaction is
regularized by the  function  $R_{\Lambda}$, for which a gaussian
form is used. 

Heavy mesons have been included in this model
 to achieve a unified description of all
mesons. Especially the parametrization of confinement should be
universal. As shown in \cite{Andre}, this is done most successfully by
extending the instanton induced residual interaction in a naive way
for heavy-light systems. This is done at the level of the interaction kernel eq. (\ref{III}) by simply allowing $f_i \in \{u,d,s,c,b\}$ in (\ref{IIImat}).
  It has to be stressed that this extension is
 purely phenomenologicaly motivated, since the derivation leading to
 (\ref{III}) is strictly valid for light quarks only. Also it should be
 mentioned that this use of the instanton induced interaction allows
 the mixing of  heavy and light flavours in the scalar and pseudoscalar
flavour-neutral sectors. This has been investigated in \cite{Andre} but will be
 neglected in this paper, since the effect turned out to be very small.

Due to the $J=0$ selection rule of the residual interaction, a gradual
fit of the model parameters to the light meson masses has been
performed as explained in \cite{Koll}, leading to the parameter sets
in table \ref{params}. The resulting spectra of the isovector mesons
are shown in fig.(\ref{isovec}). The Regge trajectory is reproduced
excellently in both models. The spectra of the light isoscalars and
kaons are of similar quality. In particular the position of $\eta$ and $\eta'$ can
be accounted for. The most significant difference between model  
$\cal A$ and $\cal B$ are the masses of the scalar mesons, {\it esp.} the $a_0(980)$ in fig.(\ref{isovec}).
  For a detailed discussion of these spectra we refer to \cite{Koll} and \cite{Ralf}. 

Based on this parameter set the  masses of the heavy quarks have been adjusted
to the $J\neq 0$ heavy mesons. Finally the additional couplings $g_{f_if_j}$
have been adjusted to reproduce the heavy-light pseudoscalar
masses. We want to stress that in this way all heavy mesons except
for the pseudoscalar $D$, $D_s$, $B$ and $B_s$ are described by
fitting the heavy quark masses only. The resulting spectra are shown
in figs. (\ref{DDs}-\ref{onium}). We find good agreement for the
heavy-light mesons in both models with small advantages in model $\cal
B$ due to a larger spin-orbit splitting. Also the gross structure of the
heavy quarkonia spectra can be reproduced, but  fine and hyperfine
splittings come out to small. This is due to the missing of a substantial
spin-orbit and spin-spin interaction, as provided {\it e.g.} by the One
Gluon Exchange potential. The radial excitations of the 
$\Upsilon$, on the other hand,  are excellently reproduced in model
$\cal B$ up to the 5s state. Thus we think that we have reached a good
overall agreement with the experimental heavy masses and therefore
have gained a good estimate for the heavy meson amplitudes. To test
these amplitudes further we investigate the semileptonic decays of heavy to light mesons. The relevant current matrix
elements are calculated in lowest order after the 
prescription of Mandelstam\cite{Mandelstam}. In our formalism this
leads in general to 
\beq
\bra{P'}J^\mu(q)\ket{P} = \intd{4}{p} \tr\left[{ \bar{\Gamma}^{P'}}({ p-\frac q2})S^F_1({\ \frac P2 +p-q}) { J^\mu} S^F_1({ \frac P2 +p}){\Gamma^P}({ p}) S^F_2({ -\frac P2 +p})\right]
\eeq
with the vertex function $ { \Gamma}^P(p)  :=  \left[S^F_1({ \frac P2 +p})\right]^{-1} { \chi}^P({ p}) \left[S^F_2({ -\frac P2 +p})\right]^{-1}$. 
In the instantaneous approximation the vertex function in the
restframe of the meson can be easily reconstructed from the Salpeter
amplitude according to 
\beqa
\Gamma^P(p)\vert_{P=(M,\vec 0)} & = &  -i \intd{3}{p'}V(\vec p, \vec p\,'){ \Phi}(\vec p\,').
\eeqa
Thus the full 4-dimensional vertex function is known in the rest frame. Formal covariance of our model then allows to calculate the vertex function of any meson on its mass shell, which is essential for the calculation of
meson decays.

\section{Semileptonic decays}

The effective Lagrangian for the semileptonic decays, {\it e.g.} $b\to c$ transitions, after integrating out the W boson, has the usual V-A current-current form
\beq
{\cal L}^{\mbox{\tiny eff}}_{cb} = -\frac{G_F}{\sqrt{2}}\,{ V_{cb}}\, {\bar c \gamma^\mu(1-\gamma^5) b}\; {\bar{\ell}\gamma_\mu(1-\gamma^5)\nu}
\eeq
with 
the Cabibbo-Kobayashi-Maskawa (CKM) matrix element $V_{cb}$. Whereas the
     matrix elements of the leptonic current can be calculated exactly, those
     of the vector ($V^\mu$) and axial vector ($A^\mu$) hadronic
     currents are parametrized by form factors, reflecting the hadronic structure. A common parametrization  is \cite{slrev}  

\vspace{2ex}

for a $0^- \to 0^-$ transition, {\it e.g.} $B \to D \ell \bar{\nu}_\ell$
\beq
\bra{D}V^\mu\ket{B}  =  f_+(q^2) (P_B^\mu +P_D^\mu)  + f_-(q^2) (P_B^\mu - P_D^\mu) 
\eeq

\vspace{2ex}

and a $0^- \to 1^-$ transition, {\it e.g.} $B \to D^* \ell \bar{\nu}_\ell$
\beqa
\bra{D^*}V^\mu\ket{B}  & = &  \frac{2i}{m_B + m_{D^*}} \varepsilon^\mu_{\;\nu\rho\sigma} \varepsilon^{*\nu}P_B^\rho P_{D^*}^\sigma  V(q^2)
\eeqa
and
\beqa
\bra{D^*}A^\mu\ket{B} 		       &  =  &   2m_{D^*} \frac{\varepsilon^* \cdot q}{q^2}q^\mu A_0(q^2) + (m_B + m_{D^*})\left(\varepsilon^{*\mu}-\frac{\varepsilon^* \cdot q}{q^2}q^\mu\right)A_1(q^2)    \nonumber\\
				       &     & -\frac{\varepsilon^* \cdot q}{m_B + m_{D^*}}\left(P_B^\mu + P_{D^*}^\mu - \frac{m_B^2-m_{D^*}^2}{q^2}q^\mu\right)A_2(q^2),
\eeqa
where $q=P_B-P_{D^{(*)}}$ is the momentum transfer and $\varepsilon$ the polarization vector of the vector meson. In the limit of vanishing lepton mass only 4 of these 6 form factors contribute to the decay rates, namely $f_+, V, A_1$ and $A_2$, since terms proportional to $q^\mu$ can be neglected. Then the differential decay rates, expressed in terms of these form factors, are
\beqa
\frac{\df\Gamma^{0^- \to 0^- }}{\df q^2} & = & |V_{cb}|^2\frac{G_F^2}{24\pi^3}P_{D}^3|f_+|^2,\\
\frac{\df\Gamma^{0^- \to 1^- }}{\df q^2} & = & |V_{cb}|^2\frac{G_F^2}{(2\pi)^3}\frac{q^2 P_{D^*}}{12m_B^2}\left(|H_+|^2+|H_-|^2+|H_0|^2\right),
\eeqa
where we have introduced the helicity amplitudes
\beqa
H_{\pm}(q^2) & := & (m_B + m_{D^*}){A_1(q^2)}\mp\frac{2m_BP_{D^*}}{m_B+m_{D^*}}{V(q^2)}\\
H_0(q^2) & := & \frac{m_B^2-m_{D^*}^2-q^2}{2m_{D^*}\sqrt{q^2}}(m_B+m_{D^*}){A_1(q^2)} -\frac{2m_B^2P_{D^*}^2}{m_{D^*}(m_B+m_{D^*})\sqrt{q^2}}{A_2(q^2)}.
\eeqa

With regard to the processes $B\to D^{(*)}\ell\bar\nu$, that is the
transition between heavy flavours, we want to summarize the
predictions of heavy-flavour symmetry. For these transitions, the
Heavy Quark Effective Theorie (HQET) provides the appropriate
framework. It is based on a systematic expansion in the inverse quark
mass and has been worked out by  Isgur and  Wise \cite{Isgur,Isgur2}. In the
limit $m_Q \to \infty$ the theory will become independent of the
heavy degrees of freedom. For the spectra of heavy hadrons this will
lead to degenerated doublets, corresponding to the two possible
alignments of the heavy quark spin. Recent experimental results show
evidence for this degeneracy in the spectrum of the $D$ mesons \cite{CLEODstp}. For
semileptonic decays HQET predicts a connection between the form
factors of $0^-\to 0^-$ and $0^-\to 1^-$ transitions. In particular, in the
infinite quark mass limit, the transition matrix elements can be expressed by a single
function only, the Isgur-Wise function $\xi$, which is normalized to
unity at maximum recoil. 

In this heavy-quark limit it is more natural to express the decay
amplitudes in terms of velocities rather than momenta and to introduce the
dimensionless variable $\omega := v_Bv_{D^{(*)}}=\left(m_B^2+m_{D^{(*)}}^2-q^2\right)/\left(2m_Bm_{D^{(*)}}\right)$ 
instead of the momentum transfer $q^2$. Therefore, a new set of form
factors is used\cite{slrev}, defined by 

\vspace{2ex}
$0^- \to 0^-$ :
\beqa
\frac{{\bra{D}V^\mu\ket{B}}}{\sqrt{m_Bm_D}} & = &  {h_+(\omega)}(v_B^\mu +v_D^\mu) +  {h_-(\omega)}(v_B^\mu - v_D^\mu);
\eeqa

$0^- \to 1^-$ :
\beqa
\frac{{\bra{D^*}V^\mu\ket{B}}}{\sqrt{m_Bm_{D^*}}} & = & i\varepsilon^\mu_{\;\nu\rho\sigma} \varepsilon^{*\nu}v_B^\rho v_{D^*}^\sigma \,{h_V(\omega)},\nonumber\\
\frac{{\bra{D^*}A^\mu\ket{B}}}{\sqrt{m_Bm_{D^*}}} & = & \varepsilon^{*\mu}(\omega + 1) \,{h_{A_1}(\omega)} - v_B^\mu\varepsilon^* \cdot v_{D^*}  \,{h_{A_2}(\omega)} - v_{D^*}^\mu\varepsilon^* \cdot v_B  \,{h_{A_3}(\omega)}.
\eeqa
with 
\beqa\label{hqlimit}
m_Q \to \infty &\quad : \quad & h_V=h_{A_1}=h_{A_3}=h_+={\xi}\\
               &   & h_{A_2}=h_-=0\nonumber
\eeqa
in the infinite quark mass limit.
The differential decay rates are given in this framework most conveniently by
\beqa
\frac{\df\Gamma^{0^- \to 0^- }}{\df \omega} & = & \frac{G_F^2}{48\pi^3}m_D^3(m_B+m_D)^2(\omega^2-1)^{3/2}|V_{cb}|^2{\cal G}^2(\omega)\\
\frac{\df\Gamma^{0^- \to 1^- }}{\df \omega} & = & \frac{G_F^2}{48\pi^3}m_{D^*}^3(m_B-m_{D^*})^2\sqrt{\omega^2-1}(\omega+1)^2\left[1+\frac{4\omega}{\omega+1}\frac{1-2\omega r_*+r_*^2}{(1-r_*)^2}\right]|V_{cb}|^2{\cal F}^2(\omega)\label{HQETGamma}
\eeqa
with $r_* = m_{D^*}/m_B$, where the two form factors $\cal G$ and
$\cal F$ are functions of $h_{\pm}$ and $h_V, h_{A_i}$
respectively. This parametrization has the advantage that for $m_Q \to
\infty$  $\cal F$ and $\cal G$ become equal and coincide with the
Isgur-Wise function $\xi$. 

In the following sections our results for semileptonic B and charmed
meson decays are compared to the available experimental data and to
the results of other models: the relativised constituent quark
model of Isgur and Scora (ISGW2 \cite{ISGW2}), the relativistic
calculation of Wirbel, Stech and Bauer in the infinite-momentum frame
(WSB \cite{WSB}) and the relativistic dispersion relation approach of
Melikhov and Stech (MS \cite{MS}). Since the experiments are not
able to extract form factors from their measurements due to missing
statistics, these are usually parametrized according to theoretical
predictions. Therefore we find it convenient to parameterize our results
in the same way, allowing a comparison with experimental data. The
accuracy of these parametrizations is always indicated.

\subsection{Semileptonic decays of $B$ mesons}\label{semilep}

\begin{figure}
\begingroup%
  \makeatletter%
  \newcommand{\GNUPLOTspecial}{%
    \@sanitize\catcode`\%=14\relax\special}%
  \setlength{\unitlength}{0.1bp}%
\begin{picture}(4679,2807)(0,0)%
\special{psfile=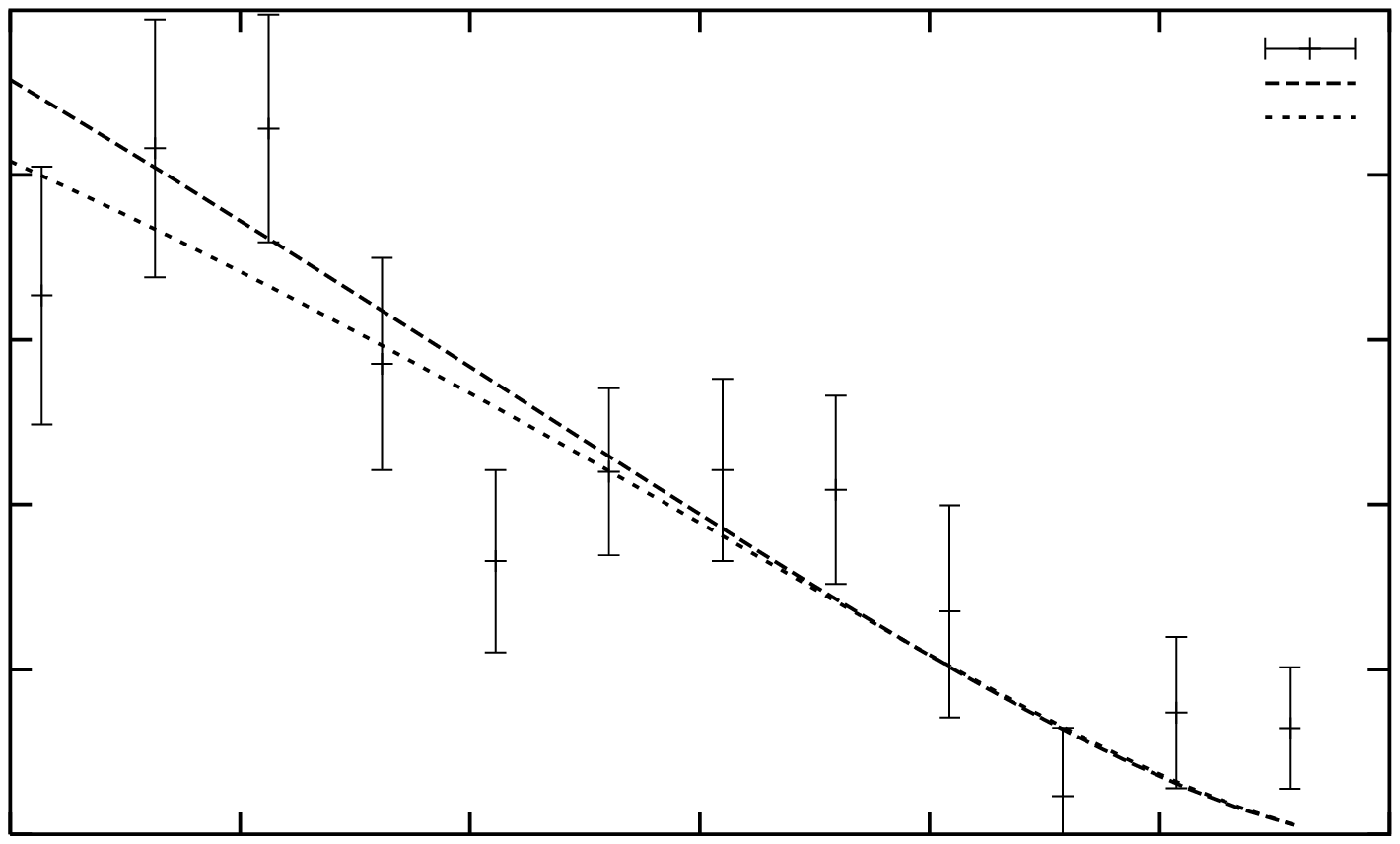 llx=0 lly=0 urx=936 ury=655 rwi=9360}
\put(4216,2494){\makebox(0,0)[r]{\tiny model $\cal B$}}%
\put(4216,2594){\makebox(0,0)[r]{\tiny model $\cal A$}}%
\put(4216,2694){\makebox(0,0)[r]{\tiny CLEO \cite{CLEOD}}}%
\put(2614,50){\makebox(0,0){$q^2 [GeV^2]$}}%
\put(100,1603){%
\special{ps: gsave currentpoint currentpoint translate
270 rotate neg exch neg exch translate}%
\makebox(0,0)[b]{\shortstack{$d \Gamma /dq^2 [ns^{-1} GeV^{-2}]$}}%
\special{ps: currentpoint grestore moveto}%
}%
\put(4629,300){\makebox(0,0){12}}%
\put(3958,300){\makebox(0,0){10}}%
\put(3286,300){\makebox(0,0){8}}%
\put(2615,300){\makebox(0,0){6}}%
\put(1943,300){\makebox(0,0){4}}%
\put(1272,300){\makebox(0,0){2}}%
\put(600,300){\makebox(0,0){0}}%
\put(550,2326){\makebox(0,0)[r]{2}}%
\put(550,1363){\makebox(0,0)[r]{1}}%
\put(550,400){\makebox(0,0)[r]{0}}%
\end{picture}%
\endgroup
\vspace*{2ex}
\caption{The differential decay rate for $B\to D\ell\bar\nu$. The data are taken from \protect\cite{CLEOD}. The values of $|V_{cb}|$ corresponding to our calculations are  $|V_{cb}| = 0.034$  and $|V_{cb}| = 0.035$ for model $\cal A$ and $\cal B$, respectively.\label{CLEO_BD} }
\end{figure}

\begin{figure}
\begingroup%
  \makeatletter%
  \newcommand{\GNUPLOTspecial}{%
    \@sanitize\catcode`\%=14\relax\special}%
  \setlength{\unitlength}{0.1bp}%
\begin{picture}(4679,2807)(0,0)%
\special{psfile=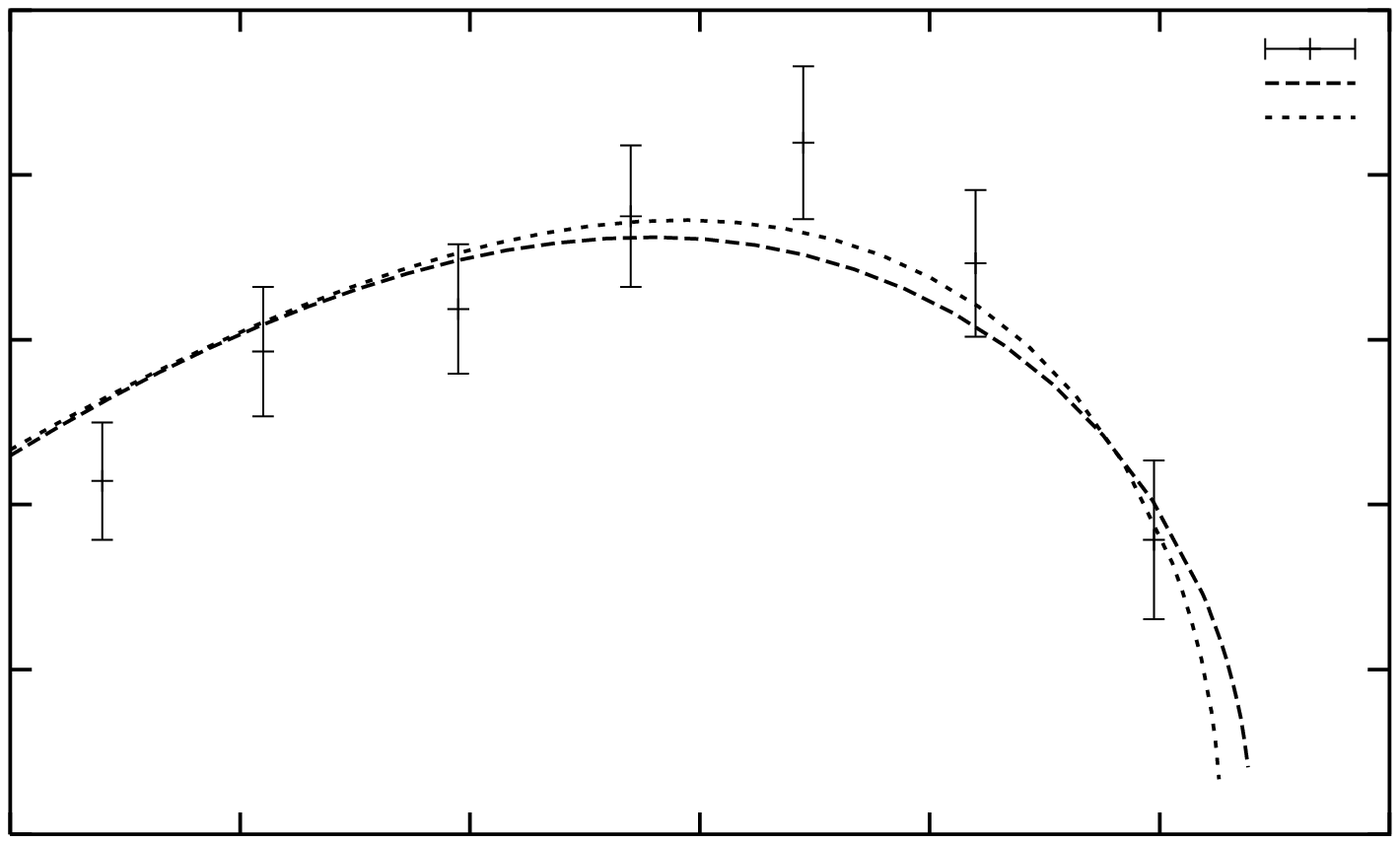 llx=0 lly=0 urx=936 ury=655 rwi=9360}
\put(4216,2494){\makebox(0,0)[r]{\tiny model $\cal B$}}%
\put(4216,2594){\makebox(0,0)[r]{\tiny model $\cal A$}}%
\put(4216,2694){\makebox(0,0)[r]{\tiny CLEO\cite{CLEODstar}}}%
\put(2614,50){\makebox(0,0){$q^2 [GeV^2]$}}%
\put(100,1603){%
\special{ps: gsave currentpoint currentpoint translate
270 rotate neg exch neg exch translate}%
\makebox(0,0)[b]{\shortstack{$d \Gamma /dq^2  [ns^{-1} GeV^{-2}]$}}%
\special{ps: currentpoint grestore moveto}%
}%
\put(4629,300){\makebox(0,0){12}}%
\put(3958,300){\makebox(0,0){10}}%
\put(3286,300){\makebox(0,0){8}}%
\put(2615,300){\makebox(0,0){6}}%
\put(1943,300){\makebox(0,0){4}}%
\put(1272,300){\makebox(0,0){2}}%
\put(600,300){\makebox(0,0){0}}%
\put(550,2807){\makebox(0,0)[r]{5}}%
\put(550,2326){\makebox(0,0)[r]{4}}%
\put(550,1844){\makebox(0,0)[r]{3}}%
\put(550,1363){\makebox(0,0)[r]{2}}%
\put(550,881){\makebox(0,0)[r]{1}}%
\put(550,400){\makebox(0,0)[r]{0}}%
\end{picture}%
\endgroup
 
\vspace*{2ex}
\caption{The differential decay rate for $B\to D^*\ell\bar\nu$. The data are taken from \protect\cite{CLEODstar}. The values of $|V_{cb}|$ corresponding to our calculations are  $|V_{cb}| = 0.034$  and $|V_{cb}| = 0.035$ for model $\cal A$ and $\cal B$, respectively.\label{CLEO_BDstar} }
\end{figure}

The decays $B \to D \ell \bar{\nu}_\ell$ and $B \to D^* \ell
\bar{\nu}_\ell$ have been measured by {\sc Cleo} \cite{CLEOD,CLEODstar} and, more recently, by {\sc Opal} \cite{OPAL} and {\sc Cleo}
\cite{CLEO2k}. Figs. (\ref{CLEO_BD}) and  (\ref{CLEO_BDstar}) show our
results for the differential decay rate compared to the  {\sc Cleo} data \cite{CLEOD,CLEODstar}.
We find good overall agreement with  the experimental data for both
decays, using a CKM matrix element of $|V_{cb}| = 0.034 \pm 0.001$ and
$|V_{cb}| = 0.035 \pm 0.001$ for model $\cal A$ and $\cal B$
respectively, which has been determined by a $\chi^2$ fit. These
values are  somewhat smaller than the PDG average of $|V_{cb}| = 0.037
\mbox{ to } 0.043$. The resulting decay rates are

\beqa
\mbox{model } {\cal A} & : & \qquad\Gamma(B\to D\ell\bar\nu_\ell) = 1.22\cdot 10^{10} s^{-1}, \qquad \Gamma(B\to D^*\ell\bar\nu_\ell) = 3.21\cdot 10^{10} s^{-1}\nonumber\\
\mbox{model } {\cal B} & : & \qquad\Gamma(B\to D\ell\bar\nu_\ell) = 1.14\cdot 10^{10} s^{-1}, \qquad \Gamma(B\to D^*\ell\bar\nu_\ell) = 3.24\cdot 10^{10} s^{-1}\nonumber.
\eeqa
The comparison with the current world average of the Particle Data Group \cite{PDG2k}

\beqa
\mbox{\sc PDG } & : & \qquad\Gamma(B^+\to \bar D^0\ell^+\nu_\ell) = 1.30 \pm 0.13\cdot 10^{10} s^{-1}, \qquad \Gamma(B^+\to\bar D^{*0}\ell^+\nu_\ell) = 3.21 \pm 0.48 \cdot 10^{10} s^{-1}\nonumber\\
	    &  & \qquad\Gamma(B^0\to D^-\ell^+\nu_\ell) = 1.36 \pm 0.12\cdot 10^{10} s^{-1}, \qquad \Gamma(B^0\to D^{*-}\ell^+\nu_\ell) = 2.97 \pm 0.17\cdot 10^{10} s^{-1}\nonumber
\eeqa
as well as the recent data of {\sc Cleo}\cite{CLEO2k} and {\sc Opal}\cite{OPAL}

\beqa
\mbox{{\sc Cleo} } & : & \qquad\Gamma(\bar B^0\to D^{*+}\ell^-\bar\nu_\ell) = 3.66 \pm 0.18 \pm 0.23 \cdot 10^{10} s^{-1}\nonumber\\
\mbox{{\sc Opal} } & : & \qquad\Gamma(\bar B^0\to D^{*+}\ell^-\bar\nu_\ell) = 3.40 \pm 0.13 \pm 0.30\cdot 10^{10} s^{-1}\nonumber
\eeqa
shows satisfying agreement. But the new data tends to higher values of $\Gamma$, which would lead to higher values of $|V_{cb}|$ from our calculation.

 In table \ref{semilepBD2} our results for
polarization ratios, defined by
\beqa
\frac{\Gamma_L}{\Gamma_T} =  \frac{\int\df q^2q^2 P_{D^*}|H_0(q^2)|^2}{\int\df q^2q^2 P_{D^*}\left(|H_+(q^2)|^2+|H_-(q^2)|^2\right)}, \qquad \frac{\Gamma_+}{\Gamma_-} =  \frac{\int\df q^2q^2 P_{D^*}|H_+(q^2)|^2}{\int\df q^2q^2 P_{D^*}|H_-(q^2)|^2},
\eeqa
 are summarized. In these ($|V_{cb}|$ independent) quantities we find good agreement
with the experimental values as well.

We also show the results for the form factor ratios
\beqa
R_1 & := & \frac{h_V(\omega)}{h_{A_1}(\omega)} =  \left(1-\frac {q^2}{(m_B+m_{D^*})^2}\right)\frac{V(q^2)}{A_1(q^2)}\\
R_2 & := & \frac{h_{A_3}(\omega)+ \frac{m_{D^*}}{m_B}h_{A_2}(\omega)}{h_{A_1}(\omega)} =  \left(1-\frac {q^2}{(m_B+m_{D^*})^2}\right)\frac{A_2(q^2)}{A_1(q^2)}.
\eeqa
In the heavy quark limit these ratios are predicted to be constant and equal to unity. In reality, due to
corrections to this limit, $R_1$ and $R_2$ do depend on $\omega$,
but at the scale of the $b$-quark mass this dependency is expected to be
very weak. Therefore these ratios are prefered in the analysis of
$B\to D^*\ell\bar\nu$, where they are assumed to be constant, whereas $h_{A_1}$ is approximated by a linear or quadratic function
\beq
h_{A_1}(\omega) \approx h_{A_1}(1)\left(1-\rho^2_{A_1}(\omega - 1)+\lambda_{A_1}(\omega - 1)^2\right)\label{ha1}.
\eeq
For our form factors the $\omega$-dependency of $R_{1/2}$ is less than 2\%, and the
mean values are shown in table \ref{semilepBD2}. The form factor $h_{A_1}$ can be described by the parametrization (\ref{ha1}) with an accuracy of 0.1\% (0.3\%) for model $\cal A$($\cal B$), yielding
\beqa
\mbox{model } {\cal A}	& : & \quad h_{A_1}(1) = 0.97	\quad	 \rho_{A_1}^2 = 0.73	\quad	 \lambda_{A_1} = 0.27\nonumber\\
\mbox{model } {\cal B} 	& : & \quad h_{A_1}(1) = 1.01	\quad	 \rho_{A_1}^2 = 0.98	\quad	 \lambda_{A_1} = 0.48,\nonumber
\eeqa
which is in good agreement with the experimental data shown in table \ref{semilepBD2}.

Two recent measurements of $B\to D^*\ell\bar\nu$ by the {\sc Cleo}
\cite{CLEO2k} and the {\sc Opal} \cite{OPAL} collaborations have been
published,  where a different parametrization of the form factor based on
dispersion relations has been used. These analyses expand $h_{A_1}$ in
the variable $z(\omega) = (\sqrt{\omega+1}-\sqrt{2})/(\sqrt{\omega+1}+\sqrt{2})$ and use the
parametrization \cite{caprini} 
$$ h_{A_1}(\omega)/h_{A_1}(1) = 1-8\rho^2z+(53\rho^2-15)z^2 - (231\rho^2-91)z^3$$
with the (inconsistent) results
$$\mbox{{\sc Opal}\cite{OPAL} : } \rho^2 = 1.21 \pm 0.12 \pm 0.20, \qquad \mbox{{\sc Cleo}\cite{CLEO2k} :
} \rho^2 = 1.67 \pm 0.11 \pm 0.22.$$ 
A similar fit to our form factor gives
$$ \mbox{model } {\cal A}	:  \rho^2 = 0.83, \qquad \mbox{model }	{\cal
B} : \rho^2 = 1.06$$
with an accuracy of 1\% and 0.4\% for model  $\cal A$ and $\cal B$, 
respectively. Thus  our calculation is rather compatible with the {\sc Opal} result.

To connect our results with the description in the framework of HQET,
we have calculated the form factors $\cal F$ and $\cal G$ (see
eq. (\ref{HQETGamma})) and their slopes  at $\omega= 1$ by fitting a quadratic function analogous to (\ref{ha1}).
This is possible with an accuracy of  0.2\% (0.3\%) for model $\cal A$($\cal B$) and gives the result
\beqa
\mbox{model } {\cal A}	& : & \quad {\cal G}(1) = 1.01,	\quad	 \rho_{\cal G}^2 = 0.70,	\quad	 \lambda_{\cal G} = 0.22;\qquad {\cal F}(1) = 0.97,	\quad	 \rho_{\cal F}^2 = 0.65,	\quad	 \lambda_{\cal F} = 0.20\nonumber\\
\mbox{model } {\cal B}	& : & \quad {\cal G}(1) = 1.01,	\quad	 \rho_{\cal G}^2 = 0.85,	\quad	 \lambda_{\cal G} = 0.31;\qquad {\cal F}(1) = 1.01,	\quad	 \rho_{\cal F}^2 = 0.91,	\quad	 \lambda_{\cal F} = 0.56\nonumber
\eeqa
These quantities ${\cal G}(1)$ and ${\cal F}(1)$ have been calculated
in the HQET. Current values are \cite{PDG2k}
$$
{\cal G}(1) = 1.00 \pm 0.07, \qquad {\cal F}(1) = 0.92 \pm 0.05.
$$

Finally we have performed the heavy quark limit numerically in our
model by scaling the quark masses with a large factor. We find that
the  form factors then indeed coincide or vanish (see eq. (\ref{hqlimit})).
 The resulting universal function $\tilde\xi$,
which we identify as the Isgur-Wise function of our model, can be
parametrized up to deviations of less than 0.2\% and 0.5\% for model $\cal A$ and $\cal B$  as $$\tilde\xi(\omega) =
\tilde\xi(1)(1-\rho^2_{\tilde\xi}(\omega-1)+\lambda_{\tilde\xi}(\omega-1)^2),$$ where 

\beqa
\mbox{model } {\cal A}	& : & \quad \tilde\xi(1) = 1.00	\quad	 \rho_{\tilde\xi}^2 = 0.78	\quad	 \lambda_{\tilde\xi} = 0.31\nonumber\\
\mbox{model } {\cal B} 	& : & \quad \tilde\xi(1) = 1.00	\quad	 \rho_{\tilde\xi}^2 = 1.06	\quad	 \lambda_{\tilde\xi} = 0.50.\nonumber
\eeqa
In particular, we find that $\tilde\xi$ is indeed normalised to $\tilde\xi(1)=1$. The $\omega$-dependence is of course model dependent.

\subsection{Semileptonic decays of charmed mesons}
The semileptonic decays of charmed mesons, induced by the quark level process 
$c\to s$, have been measured for the $D\to K^{(*)}$ as well as for the $D_s\to\eta/\eta'/\phi$ transitions.
The results, averaged over isospin, are shown in table \ref{DK} and \ref{Dseta}.
For the $D$ meson decay $D\to K\ell\bar\nu$ the comparison with our
calculation shows reasonable agreement for both models. 
The decay to the $K^*$ final state however
is overestimated by about a factor of 2. The polarization observables
on the other hand are comparable with the experimental result,
whereby model $\cal A$ gives better agreement than model $\cal
B$. This is a well known problem of constituent quark models. Our
results are rather comparable with those of Wirbel {\it et al.} (WSB \cite{WSB}). With
respect to the ISGW2\cite{ISGW2} predictions it is interesting to note that, whereas the inclusion of relativistic
corrections was one of the main incredients in their model to decrease the
$D\to K^{*}$ decay rate, this problem still exists in our full relativistic
calculation.

Apart from the decay rates and the polarization observables, the form factor ratios at zero momentum transfer, defined by
\beq
r_V  := \frac {V(0)}{A_1(0)}, \qquad\qquad r_2  := \frac {A_2(0)}{A_1(0)},
\eeq
 are considered. To extract these ratios from experiment the form factors
are usually parametrized by a  simple pole ansatz  
with a pole mass of 2.1 GeV for the vector form factor and 2.5 GeV 
for the axial form factors. We have performed such a fit to our calculations, 
which works up to deviations of about 6\% for the
form factors and of 4\% for their ratios, and have extracted
the form factor ratios for $D$ meson decays from these
parametrizations. The results in table \ref{DK} show that the axial
form factors are generally overrated, while our vector form factor
parameters agree with the values extracted from experiment. Thus the
failure in the $D\to K^{(*)}$ decay rate can be traced back to this
overestimation of the axial form factors.

In this connection it is interesting to estimate the effect of the
full relativistic treatment. To do this we neglect the negative energy
or 'lower' components of the Salpeter amplitude when reconstructing
the vertex function. This procedure changes
the norm of the Salpeter amplitude by a factor of
$$
\frac{\langle \Phi^{++}|\Phi^{++}\rangle}{\langle \Phi |\Phi\rangle} = 0.55
$$
for the $D$ meson in model $\cal A$. Here  $\Phi^{++} :=
\frac{1+\gamma^0}2\Phi\frac{1-\gamma^0}2$ and  $\langle . |.\rangle$
is the scalar product induced by the Salpeter equation.
This shows the importance of  the negative energy components for heavy-light mesons.
With this reduced vertex function the form
factor ratios $r_V$ and $r_2$  are  calculated. Hereby we concentrate on model $\cal A$,
since, as has been discussed in \cite{Ralf}, the positive and negative
energy components of the Salpeter amplitude decouple in the
non-relativistic reduction of the Salpeter equation. We find that
these ratios slightly rize: $r_V$
changes from $1.48$ to $1.55$, $r_2$ goes from $0.78$ up to
$0.84$\footnote{These are the calulated values, not to results 
of the pole ansatz parametrization in table \ref{DK}}, while the $q^2$
dependence is hardly changed. Thus the modification of a
form factors by the full relativistic treatment seems to be more
complex than the intuitive correction anticipated in \cite{ISGW2}.

The results on the semileptonic decays $D_s\to \eta$ and $D_s\to
\eta'$ are shown in table \ref{Dseta}. Here it has to be
stressed that the flavour mixing of $\eta$ and $\eta'$ had already been fixed
by the mass fit. No additional mixing parameter is necessary. Although the differences between the
results of our two models are rather large, the experimental data do
not allow to prefer one of our parameter sets. 

Our results on $D_s\to \phi$ show the same behaviour as those on $D\to K^*$:
While the decay rate is overestimated by a factor of 2, the
polarization observable is in good agreement with the experimental
data and the form factor ratio $r_V$ tends to be too small. Concerning the
ratio of axial form factors it is interesting to note that the
experimental value of $r_2$ is 4 standard deviations higher than the
corresponding value for $D\to K^*$, which is in contradiction to an
(approximate) flavour $SU(3)$ symmetry. This result, if it should be
confirmed, is clearly a challenge for any theoretical description.

Finally, to conclude the discussion of charmed meson decays, we report
our results on the Cabibbo suppressed process $c\to d$ in tables
\ref{Dpi}-\ref{DsK}. The form factors again have been parametrized
by a pole ansatz with pole masses of 2.0 GeV and 2.4 GeV for the
vector and axial form factors respectively, which works up to
deviations of $8\%$ for $A_1(0)$ and $5\%$ for the ratios. The current experimental situation however allows no 
evaluation of our description of these decays.

Thus in summary we find excellent agreement in the description of
heavy to heavy transitions $B \to D^{(*)}$ over the whole kinematic
regime. We find also consistency with the description of these
processes in the framework of the HQET. The results on heavy to light transitions
are mostly in agreement with the experimental data, but the common
problem of quark models to overestimate the axial form factors is also
present here.

\section{Non-leptonic weak decays}
To extract further information from our meson amplitudes we finally
investigate non-leptonic decays. On tree level, non-leptonic decays are mediated by a single W-boson
emission. Hard and soft gluonic effects however might play a significant
role in these processes. These corrections have been calculated with
great effort in the last years in order to extract model independent
Cabibbo-Kobayashi-Maskawa matrix elements or signatures for CP
violation in $B$ decays from experimental data. Thereby the hard and
soft gluon contributions are separated by a Wilson operator product
expansion, which results in the effective Lagrangian, {\it e.g.} for $B\to
D\pi$ transitions, 
\beqa
{\cal L}^{\mbox{\tiny eff}}_{B\to D\pi} & = & -\frac{G_F}{\sqrt{2}}\,{ V_{cb}V^*_{du}}\, \left(\frac12\left(a_1(\mu) (\bar cb)^\mu_{V-A}{(\bar du)_\mu}_{V-A} + a_2(\mu) (\bar db)^\mu_{V-A}{(\bar cu)_\mu}_{V-A} \right)\right.\nonumber\\
 & &  \left.+ \frac12 C_1(\mu) (\bar d \lambda^A b)^\mu_{V-A}{(\bar c \lambda^A u)_\mu}_{V-A} + \frac12 C_2(\mu) (\bar c \lambda^A b)^\mu_{V-A}{(\bar d \lambda^A u)_\mu}_{V-A}\right)
\eeqa
with $(\bar c \lambda^A b)^\mu_{V-A} := \bar c
\gamma^\mu(1-\gamma^5)\lambda^A b$ etc., $a_1 := C_1 + \frac13 C_2$
and $a_2 := C_2 + \frac13 C_1$, where $C_{1/2}$ are the (scale
dependent) Wilson coefficients, and $\lambda^A$ the SU(3) color
Gell-Mann matrices. Here we found it convenient to use an expression
symmetric under Fierz transformation. The second line obviously does
not contribute for color singlet states. 

This lagrangian gives rise to W emission diagrams.
Contributions by weak annihilation and internal W exchange, which are
suppressed by powers of $\Lambda_{QCD}/m_b$, are neglected. 

Thus the matrix element of a product of currents has to be
evaluated. This is usually done using the "factorization
approximation", where one assumes that the amplitude is dominated by
its factorizable part. Then it is given by the product of two current
matrix elements, {\it e.g.} for the transition $B^+\to \overline{D}^0 \pi^+$ 
\beqa
A(B^+\to \overline{D}^0\pi^+) & = \;  \frac{G_F}{\sqrt{2}}V_{cb}V_{du}^*\, &\left\{a_1{\bra{\pi^+}{h_\mu}_{du}\ket{0}}{\bra{\overline{D}^0}h_{cb}^\mu\ket{B^+}} + a_2{\bra{\overline{D}^0}{h_\mu}_{dc}\ket{0}}{\bra{\pi^+}h_{ub}^\mu\ket{B^+}}\right\}.
\eeqa

In this way the decay amplitude can be expressed by the decay constant and a form factor of the semileptonic decay at the relevant $q^2$, {\it e.g.}
\beqa
{\bra{\pi^+}J^\mu\ket{0}}{\bra{\overline{D}^0}J'_\mu\ket{B^+}} & = & \left(m_B^2-m_D^2\right){f_\pi} {F_0(m_\pi^2)}\\
{\bra{\pi^+}J^\mu\ket{0}}{\bra{\overline{D^*}^0}J'_\mu\ket{B^+}} & = & 2 \varepsilon^* \cdot p_B m_{D^*}{f_\pi} {A_0(m_\pi^2)}
\eeqa
with the decay constants $\bra{0}J^\mu\ket{0^-}  =  i { f } P^\mu$, $
\bra{0}J^\mu\ket{1^-} = m { F } \varepsilon^\mu$.  It should be noted that
the relevant form factors are $F_0$ and $A_0$, which are unimportant
for semileptonic decays due to the smallness of the lepton
mass. Hence non-leptonic decays offer a possibility to access the
remaining semileptonic form factors, at least within the framework of
the factorization approximation. The  factorization assumption has
been proven recently at two loop order \cite{Beneke} for a special
class of decays. 

In this approach, however, the resulting amplitudes are scale dependent
due to the $\mu$-dependence of the Wilson coefficients, which is not
canceled by the scale independent matrix elements. To deal with this
difficulty the coefficients $a_1$ and $a_2$ are often treated as
effective  and free parameters, corresponding to some factorization
scale,  to be extracted from
the data. But since we are interested in an estimate of the quality
of our form factors, we find it sufficient to neglect all strong
gluonic effects, which results in $C_1 = 1, C_2 = 0 \equiv a_1 = 1,
a_2 = \frac13$ and restrict our calculation to the decays of $B$ mesons
via the emission of a W meson,  usually called Type I decays. 

The weak decay constants are shown in table \ref{weakdec}. Since these are generally overestimated in our models we
have used the experimental values, which are extracted from leptonic
decay or $\tau-$decay and are summarized in table \ref{weakdec}. For
the vector mesons $D^*$ and $D_s^*$, where no data are available yet,
we use $F_{D_{(s)}^*} \approx f_{D_{(s)}}$, which is valid in the
heavy quark limit. Deviations from this limit are expected to be about 10-20\%\cite{Neubert}. 
The relevant CKM-matrix elements are taken from \cite{PDG2k}
except for $V_{cb}$ where we use the fit results from section
\ref{semilep}. 

Our results for non-leptonic $B$-decays are shown in table
\ref{nonlept}, compared with the experimental data from \cite{PDG2k}
as well as the calculation of Neubert {\it et al.} \cite{Neubert}. 
We find good agreement with the data
available so far for both our models; and for
those decays, which are not measured yet, our results are comparable
with \cite{Neubert}.

In order to stress the influence of  our form factors we show the
ratios of decay rates (in which the decay constants as well as the
coefficient $a_1$ cancel) in  table \ref{nonlept_ratios}. We also find
good agreement with the experimental data, however the errors are
quite large.

\section{Summary}

In this paper we have calculated the masses and the exclusive semileptonic and
non-leptonic decays of heavy mesons in a constituent quark model
based on the Bethe-Salpeter equation in instantaneous
approximation. Apart from a linear confinement potential with two
different phenomenological Dirac structures, a flavour dependent
residual interaction motivated by instanton effects is adopted, which has been naively generalized to
heavy flavours. Thus extending a very good description of light mesons,
which has recently been updated in \cite{Koll}, we also find good
overall agreement with the data on heavy meson masses. 

The resulting
meson amplitudes are used to calculate semileptonic decay rates. We find
excellent agreement in the description of heavy to heavy transitions $B
\to D^{(*)}$ over the whole kinematic regime. Our results are also
consistent with the description of these processes in the framework of
the HQET. The results on heavy to light transitions
are mostly in agreement with the experimental data, but the common
problem of quark models to overestimate the axial form factors is
also found here.

Finally non-leptonic decays have then been calculated in the
approximation of factorizing matrix elements. In  spite of this simple
picture we find good agreement with the experimental results on $B$
meson decays.

\noindent {\bf Acknowledgement:} Financial support by funds provided
by the Graduiertenkolleg ``Die Erforschung subnuklearer Strukturen der Materie'' is gratefully acknowledged.

\newpage
\section*{FIGURES}
\vspace*{-0.5cm}
\begin{center}
\begin{figure}
\input{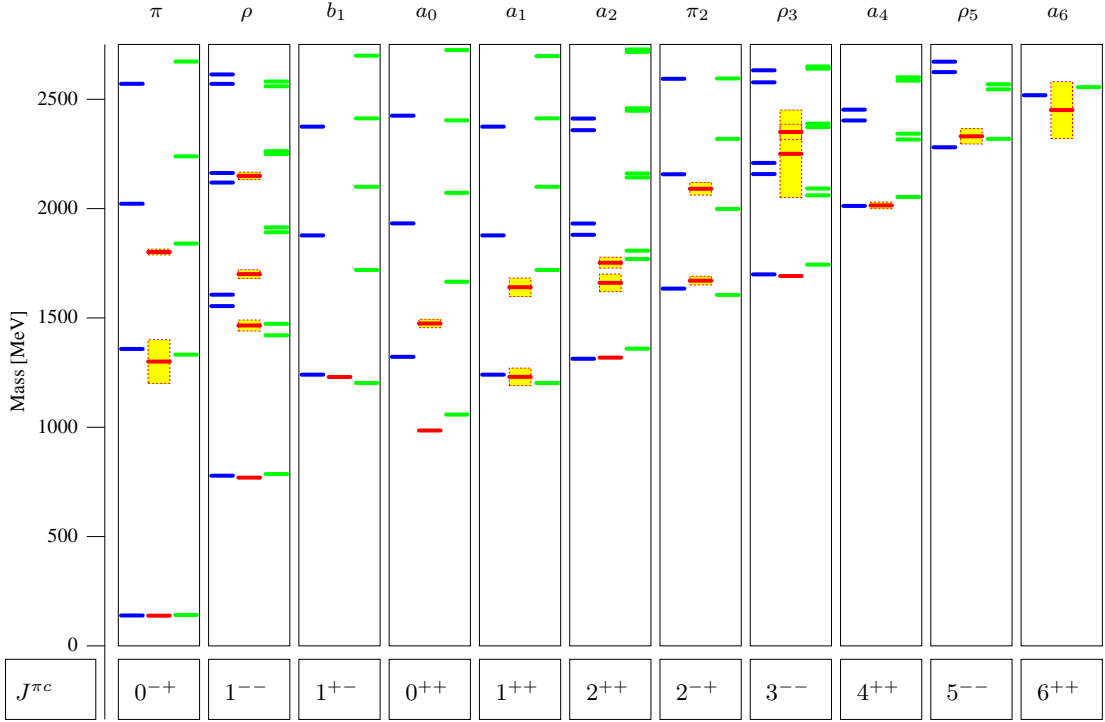}\\[1ex] 
\caption{The spectrum of the isovector mesons. The centric column shows the experimental data from \protect\cite{PDG2k}. Errors are indicated by  shadowed boxes. The left and right columns show our results with model $\cal A$ and $\cal B$, respectively. }\label{isovec}
\end{figure}
\begin{figure}
\hfill
\input{Weak_Decays_D.pstex_t}
\hfill
\input{Weak_Decays_Ds.pstex_t}
\hspace*{1cm}\\[1ex]
\caption{The spectra of the charmed $D$ and $D_s$ mesons.  The centric column shows the experimental data from \protect\cite{PDG2k}. Errors are indicated by  shadowed boxes. The levels marked with {\tiny $^\dagger$} are taken from \protect\cite{CLEODstp}. The left and right columns show our results with model $\cal A$ and $\cal B$, respectively. Note that the $1^+$ states are twofold degenerated in our calculation corresponding to the  total spin $S=0$ and $S=1$.}\label{DDs}
\end{figure}

\begin{figure}
\hfill
\input{Weak_Decays_B.pstex_t} 
\hfill
\input{Weak_Decays_Bs.pstex_t} 
\hspace*{1cm}\\[2ex]
\caption{The spectra of the bottom $B$ and $B_s$ mesons.  The centric column shows the experimental data from \protect\cite{PDG2k}. Errors are indicated by  shadowed boxes. The levels marked with {\tiny $^\dagger$} are taken from \protect\cite{Ciulli}. The left and right columns show our results with model $\cal A$ and $\cal B$, respectively. Note that the $1^+$ states are twofold degenerated in our calculation corresponding to the  total spin $S=0$ and $S=1$.}\label{BBs}
\end{figure}

\begin{figure}
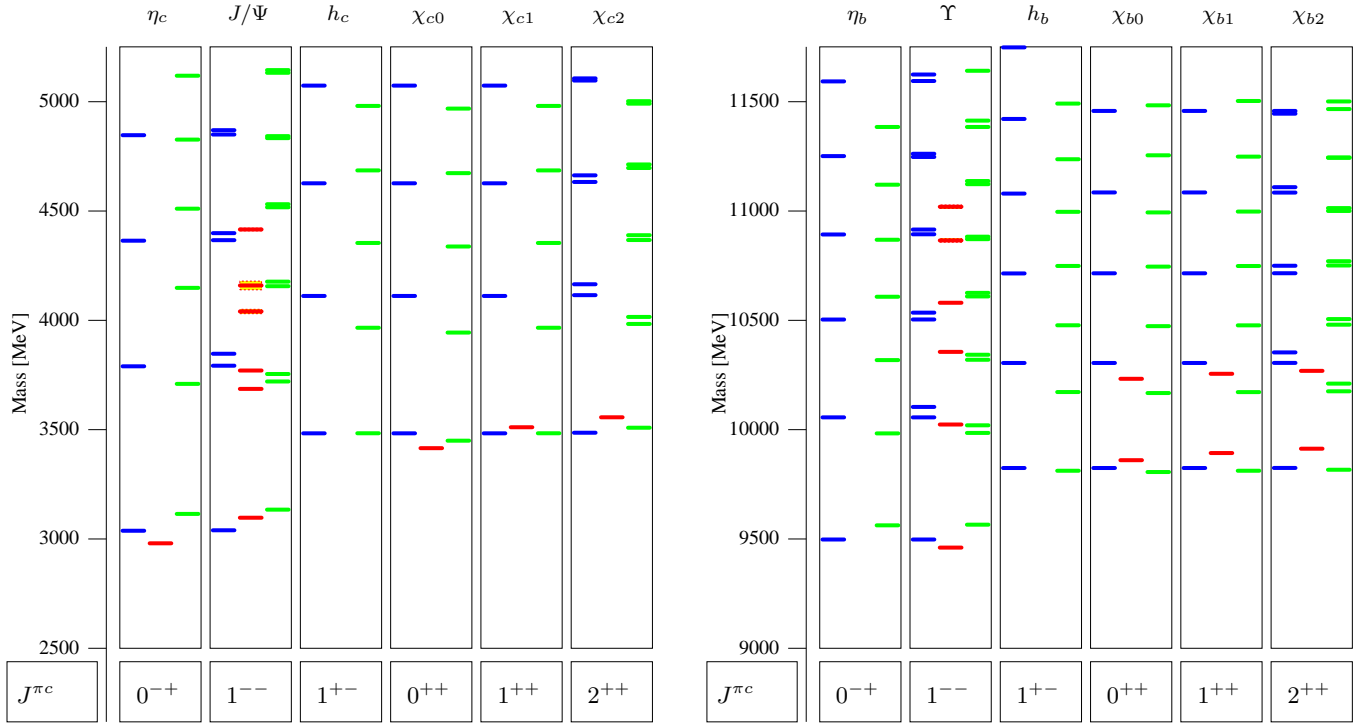

\input{Weak_Decays_Charmonium.pstex_t} 
\hfill
\input{Weak_Decays_Bottomonium.pstex_t}\\[2ex] 
\caption{The spectra of heavy quarkonia. The centric column shows the experimental data from \protect\cite{PDG2k}. Errors are indicated by  shadowed boxes. The left and right columns show our results with model $\cal A$ and $\cal B$, respectively. }\label{onium}
\end{figure}
\end{center}

\section*{TABLES}

\table
\begin{table}
\begin{tabular}{cldd}
\\
 		& \bf Parameter				& \bf Model $\cal A$	&  \bf Model $\cal B$       \\
\\\hline\\
 		&$m_n$ \space [MeV] 			& 306 			& 380                   \\[0,5ex]
\sl Constituent	&$m_s$ \space [MeV] 			& 503 			& 550                   \\[0,5ex]
\sl quark masses	&$m_c$ \space [MeV]		& 1835			& 1780                  \\[0,5ex]
		&$m_b$ \space [MeV]			& 5240			& 5092                  \\
\\
		&$g_{nn}$ \space [GeV$^{-2}$]		& 1.73	 		& 1.62                 \\[0,5ex]
		&$g_{ns}$ \space [GeV$^{-2}$]		& 1.54 			& 1.35                 \\[0,5ex]
\sl Residual	&$g_{nc}$ \space [GeV$^{-2}$]      	& 1.11 			& 1.58                \\[0,5ex]
\sl   interaction	&$g_{nb}$ \space [GeV$^{-2}$]   & 0.53  		& 1.07                 \\[0,5ex]
	 	&$g_{sc}$ \space [GeV$^{-2}$]      	& 0.65  		& 1.27                 \\[0,5ex]
		&$g_{sb}$ \space [GeV$^{-2}$]      	& 0.00			& 0.76                 \\[0,5ex]
		&$\Lambda$ \space [fm] 			& 0.30			& 0.42                  \\ 
\\
\sl Confinement	&$a_c$ \space [GeV]			& $-$1.751 		& $-$1.135              \\[0,5ex]
\sl parameters	&$b_c$ \space [GeV fm$^{-1}$]     	&   2.076		& 1.300                 \\
\\
\sl Spin structure	&$\mit\Gamma\otimes\mit\Gamma$	& $\frac12\left(\id\otimes\id - \gamma^0\otimes\gamma^0\right)$	&  $\frac12\left(\id\otimes\id - \gamma^5\otimes\gamma^5-\gamma_\mu\otimes\gamma^\mu\right)$	\\
\\
\end{tabular}
\vspace*{0.5ex}
\caption{The parameters of the confinement force, the 't Hooft interaction and the constituent quark masses in the models $\cal A$ and $\cal B$.\label{params}}
\end{table}
\endtable
\table
\renewcommand{\arraystretch}{1}
\begin{tabular}{lcdd@{\hspace{2em}}|lcdd}
&&&&&&&\\
\bf Meson ($J^{\pi}$)	& \bf  n	& \bf Model $\cal A$	& \bf Model $\cal B$	& \bf Meson ($J^{\pi}$)	&  \bf n	& \bf Model $\cal A$	& \bf Model $\cal B$\\
&&&&&&&\\\hline&&&&&&&\\
$D(0^-)$		&  0	& 1869			& 1869			& $D_s(0^-)$		&  0	& 1969			& 1969\\
			&  1	& 2677			& 2578			& 			&  1	& 2794			& 2683\\
			&  2	& 3258			& 3041			&			&  2	& 3388			& 3152\\
&&&&&&&\\												    
$D^*(1^-)$		&  0	& 1993			& 2034			& $D_s^*(1^-)$		&  0	& 2049			& 2116\\
			&  1	& 2769			& 2648			&			&  1	& 2852			& 2746\\
			&  2	& 2822			& 2679			& 			&  2	& 2905			& 2776\\
			&  3    & 3327			& 3079			&			&  3	& 3432			& 3192\\
			&  4    & 3344			& 3101			&			&  4	& 3454			& 3210\\
&&&&&&&\\
$D_0^*(0^+)$		&  0	& 2519			& 2375			& $D_{s0}^*(0^+)$	&  0    & 2563			& 2464\\
			&  1	& 3115			& 2884			&            		&  1    & 3196			& 2986\\
&&&&&&&\\												        
$D_1(1^+)$		&  0	& 2464			& 2420			& $D_{s1}(1^+)$		&  0    & 2532			& 2506\\
			&  1	& 2464	 		& 2420			&   			&  1    & 2532			& 2506\\
			&  2	& 3075			& 2908			&            		&  2    & 3172			& 3007\\
			&  3	& 3075			& 2908 			&            		&  3    & 3172			& 3007\\
&&&&&&&\\												        
$D_2^*(2^+)$		&  0	& 2475			& 2469			& $D_{s2}^*(2^+)$      	&  0    & 2541			& 2552\\
			&  1	& 3086			& 2930			& 			&  1    & 3182			& 3032\\
			&  2	& 3128			& 2957			&			&  2    & 3223			& 3057\\
&&&&&&&\\
\end{tabular}
\vspace*{0.5ex}
\caption{Masses of the charmed $D$ and $D_s$ mesons in MeV, calculated in model $\cal A$ and $\cal B$; n denotes the radial excitation.  Note that the $1^+$ states are twofold degenerated  corresponding to the  total spin $S=0$ and $S=1$.}

\vfill
\renewcommand{\arraystretch}{1}
\begin{tabular}{lcdd@{\hspace{2em}}|lcdd}
&&&&&&&\\
\bf Meson ($J^{\pi}$)	& \bf  n	& \bf Model $\cal A$	& \bf Model $\cal B$	& \bf Meson ($J^{\pi}$)	&  \bf n	& \bf Model $\cal A$	& \bf Model $\cal B$\\
&&&&&&&\\\hline&&&&&&&\\
$B(0^-)$		&  0	& 5279			& 5279			& $B_s(0^-)$		&  0	& 5368			& 5369\\
			&  1	& 6002			& 5869			& 			&  1	& 6101			& 5960\\
			&  2	& 6517			& 6279			&			&  2	& 6628			& 6376\\
&&&&&&&\\			  			      						    	  			  
$B^*(1^-)$		&  0	& 5325			& 5346			& $B_s^*(1^-)$		&  0	& 5369			& 5425\\
			&  1	& 6035			& 5900			&			&  1	& 6102			& 5986\\
			&  2	& 6108			& 5947			& 			&  2	& 6172			& 6031\\
			&  3    & 6543			& 6296			&			&  3	& 6629			& 6392\\
			&  4    & 6575			& 6322			&			&  4	& 6664			& 6417\\
&&&&&&&\\			  			      							  			  
$B_0^*(0^+)$		&  0	& 5796			& 5675			& $B_{s0}^*(0^+)$	&  0    & 5822			& 5750\\
			&  1	& 6341			& 6120			&            		&  1    & 6401			& 6208\\
&&&&&&&\\			  			      						          			  
$B_1(1^+)$		&  0	& 5770			& 5696			& $B_{s1}(1^+)$         &  0    & 5822			& 5774\\
			&  1	& 5770	 		& 5696			&   			&  1    & 5822			& 5774\\
			&  2	& 6322			& 6136			&            		&  2    & 6401			& 6224\\
			&  3	& 6322			& 6136 			&            		&  3    & 6401			& 6224\\
&&&&&&&\\			  			      						          			  
$B_2^*(2^+)$		&  0	& 5771			& 5711			& $B_{s2}^*(2^+)$	&  0    & 5823			& 5788\\
			&  1	& 6324			& 6140			& 			&  1    & 6402			& 6230\\
			&  2	& 6391			& 6186			&			&  2    & 6466			& 6273\\
&&&&&&&\\
\end{tabular}
\vspace*{0.5ex}
\caption{Masses of the bottom $B$ and $B_s$ mesons in MeV, calculated in model $\cal A$ and $\cal B$; n denotes the radial excitation.  Note that the $1^+$ states are twofold degenerated  corresponding to the  total spin $S=0$ and $S=1$.}
\endtable

\clearpage
\table
\renewcommand{\arraystretch}{1}
\begin{tabular}{lrdd@{\hspace{2em}}|lcdd}
&&&&&&&\\
\bf Meson ($J^{\pi c}$)	& \bf  n	& \bf Model $\cal A$	& \bf Model $\cal B$	& \bf Meson ($J^{\pi c}$)	&  \bf n	& \bf Model $\cal A$	& \bf Model $\cal B$\\
&&&&&&&\\\hline&&&&&&&\\
$\eta_c(0^{-+})$	&  0	& 3037			& 3114			& $\eta_b(0^{-+})$       &  0	& 9497			& 9562\\
			&  1	& 3789			& 3708			&         		 &  1	& 10055			& 9983\\
			&  2	& 4363			& 4147			&                        &  2	& 10503			& 10318\\
&&&&&&&\\
$J/\Psi(1^{--})$	&  0	& 3039			& 3133			& $\Upsilon(1^{--})$     &  0	& 9497			&  9565\\
			&  1	& 3792			& 3719			&                        &  1	& 10055			&  9985\\
			&  2	& 3846			& 3754			&     			 &  2	& 10103			& 10020\\
			&  3    & 4366			& 4155			&                        &  3	& 10503			& 10319\\
			&  4    & 4398			& 4176			&                        &  4	& 10534			& 10342\\
			&      	& 			& 			&                        &  5	& 10892			& 10609\\
			&      	& 			& 			&                        &  6	& 10914			& 10625\\
			&      	& 			& 			&                        &  7	& 11250			& 10871\\
			&      	& 			& 			&                        &  8	& 11261			& 10882\\
&&&&&&&\\																 
$h_c(1^{+-})$		&  0	& 3482			& 3483			& $h_b(1^{+-})$          &  0   & 9824			& 9811\\
			&  1	& 4110			& 3965			&                        &  1   & 10304			& 10171\\
&&&&&&&\\
$\chi_{c0}(0^{++})$	&  0	& 3482			& 3449			& $\chi_{b0}(0^{++})$    &  0   & 9824			& 9805\\
			&  1	& 4110			& 3943			&                        &  1   & 10304			& 10166\\
&&&&&&&\\
$\chi_{c1}(1^{++})$	&  0	& 3482			& 3483	  		& $\chi_{b1}(1^{++})$    &  0   & 9824			& 9811\\
			&  1	& 4110	 		& 3965			&			 &  1   & 10304			& 10171\\
&&&&&&&\\
$\chi_{c2}(2^{++})$	&  0	& 3485			& 3508			& $\chi_{b2}(2^{++})$    &  0   & 9824			& 9816\\
			&  1	& 4114			& 3983			&		         &  1   & 10304			& 10175\\
			&  2	& 4164			& 4015			&                        &  2   & 10352			& 10210\\
&&&&&&&\\
\end{tabular}
\vspace*{0.5ex}
\caption{Masses of the heavy quarkonia  in MeV, calculated in model $\cal A$ and $\cal B$; n denotes the radial excitation.}
\endtable

\renewcommand{\arraystretch}{1}
\table
\begin{table}
\renewcommand{\arraystretch}{1}
\begin{tabular}{lcddddd}
&&&&&&\\
\bf Parameter			& \bf Exp. \cite{PDG2k}	&\bf Model $\cal A$	& \bf Model $\cal B$	& \bf ISGW2 \cite{ISGW2}& \bf WSB \cite{WSB} &\bf MS\cite{MS}\\
&&&&&&\\
\hline
&&&&&&\\
$\Gamma (B\to D)\; \left[10^{13} |V_{cb}|^2 s^{-1}\right]$	& 	---			& 1.05	& 0.93  & 1.19	& 0.808 & 0.86 \\[0,5 ex]
$\Gamma (B\to D^*)\; \left[10^{13} |V_{cb}|^2 s^{-1}\right]$	& 		---		& 2.78	& 2.64  & 2.48	& 2.19  & 2.28  \\[0,5ex]
$\Gamma_L/\Gamma_T$						& $1.24 \pm 0.16$\cite{CLEOAvery}& 1.14	& 1.20  & 1.04 	& ---   & 1.11  \\[0,5ex]
$\Gamma_+/\Gamma_-$						&	---			& 0.23  & 0.27  & ---	& ---   &  ---      \\
&&&&&&\\
$R_1$						& $1.18 \pm 0.30 \pm 0.12$	& 1.18	& 1.10  & 1.27	& 1.09$^{*}$  &  ---  \\[0,5ex]
$R_2$						& $0.71 \pm 0.22 \pm 0.07$      & 0.94 	& 0.87 & 1.02	& 1.06$^{*}$  &  ---  \\[0,5ex]
$\rho_{A_1}^2$					& $0.91 \pm 0.15 \pm 0.06$      & 0.75	& 1.02 & ---		& ---   &  ---  \\
&&&&&&\\
\end{tabular}
\hfill {\tiny * taken from \cite{slrev}}
\caption{$B\to D^{(*)}\ell\bar\nu$ decay observables and form factor
parameters. We use $|V_{cb}| = 0.034 ({\cal A})$ resp. $|V_{cb}| =
0.035 ({\cal B})$, as described in section \ref{semilep}.}\label{semilepBD2}

\end{table}

\begin{table}
\begin{tabular}{lcddddd}
&&&&&&\\
\bf Parameter			& \bf Exp. \cite{PDG2k}	&\bf Model $\cal A$	& \bf Model $\cal B$	& \bf ISGW2 \cite{ISGW2}& \bf WSB \cite{WSB} &\bf MS\cite{MS}\\
&&&&&&\\
\hline
&&&&&&\\
$\Gamma (D\to K) \; \left[10^{10}s^{-1}\right]$		& $7.97 \pm 0.36$& 7.51	& 7.26        & 10.0	& 8.26 	& 9.7\\[0,5 ex]
$\Gamma (D\to K^*) \; \left[10^{10}s^{-1}\right]$	& $4.55 \pm 0.34$& 7.64	& 10.08       & 5.4	& 9.53 	& 6.0\\[0,5 ex]
$\Gamma_L/\Gamma_T$					& $1.14 \pm 0.08$& 1.29& 1.48        & 0.94	& 0.91 	& 1.28\\[0,5 ex]
$\Gamma_+/\Gamma_-$					& $0.21 \pm 0.04$& 0.23& 0.34        & ---	& ---	& ---\\
&&&&&&\\
$A_1(0)$				& $0.56 \pm 0.04$ \cite{slrev}   & 0.69 & 0.81		&  ---	& 0.88	& 0.66\\[0,5 ex]
$r_V$							& $1.82 \pm 0.09$& 1.54  & 1.18        	& 2.0$^{*}$& 1.44& 1.56\\[0,5 ex]
$r_2$							& $0.78 \pm 0.07$& 0.81 & 0.62         & 1.3$^{*}$& 1.31& 0.74\\
&&&&&&\\
\end{tabular}
\hfill {\tiny * taken from \cite{E791DK}}
\caption{$D\to K^{(*)}\ell\bar\nu$ decay observables and form factor
parameters. The experimental decay rates are averaged over isospin. We use $|V_{cs}| = 0.975$ \protect\cite{PDG2k}. \label{DK}}
\end{table}

\begin{table}
\begin{tabular}{lcddddd}
&&&&&&\\
\bf Parameter			& \bf Exp. \cite{PDG2k}	&\bf Model $\cal A$	& \bf Model $\cal B$	& \bf ISGW2 \cite{ISGW2}& \bf WSB \cite{WSB} &\bf MS\cite{MS}\\
&&&&&&\\
\hline
&&&&&&\\
$\Gamma (D_s\to \eta) \; \left[10^{10}s^{-1}\right]$		& $5.24 \pm 1.41$& 4.05 & 3.11       & 3.5 $^{ a)}$	& --- 	& 5.0\\[0,5 ex]
$\Gamma (D_s\to \eta') \; \left[10^{10}s^{-1}\right]$		& $1.80 \pm 0.69$& 1.27 & 1.75       & 3.0 $^{ a)}$	& --- 	& 1.85\\[0,5 ex]
$\Gamma (D_s\to \phi) \; \left[10^{10}s^{-1}\right]$		& $4.03 \pm 1.01$& 7.89 & 9.67       & 4.6	& --- 	& 5.1\\[0,5 ex]
$\Gamma_L/\Gamma_T$						& $0.72 \pm 0.18$& 1.20 & 1.42    & 0.96	& ---	& --- \\[0,5 ex]
$\Gamma_+/\Gamma_-$						& 	---	 & 0.20 & 0.33       & ---	& ---	& ---\\
&&&&&&\\
$A_1(0)$							&	---	 & 0.66  & 0.79		&  ---		& ---	& 0.65\\[0,5 ex]
$r_V$								& $1.92 \pm 0.32$& 1.77  & 1.30       	& 2.1$^{*}$	& --- 	& 1.71\\[0,5 ex]
$r_2$								& $1.60 \pm 0.24$  & 0.85& 0.63     	& 1.3$^{*}$	& --- 	& 0.72\\
&&&&&&\\
\end{tabular}
{\hfill \tiny * taken from \cite{E791Dsphi}}
\caption{$D_s\to \eta/\eta'/\phi\ell\bar\nu$ decay observables and form factor parameters. The experimental decay rates are averaged over isospin. We use $|V_{cs}| = 0.975$ \protect\cite{PDG2k}.\\ $^{a)}$ A $\eta-\eta'$ mixing angle of $-20^\circ$ is
assumed. An angle of $-10^\circ$ would lead to 5.3 and 2.3.\label{Dseta}}
\end{table}
\begin{table}
\begin{tabular}{lcddddd}
&&&&&&\\
\bf Parameter			& \bf Exp. \cite{PDG2k}	&\bf Model $\cal A$	& \bf Model $\cal B$	& \bf ISGW2 \cite{ISGW2}& \bf WSB \cite{WSB} &\bf MS\cite{MS}\\
&&&&&&\\
\hline
&&&&&&\\
$\Gamma (D^+\to \pi^0) \; \left[10^{9 }s^{-1}\right]$	& $2.9 \pm 1.4$ & 1.03  & 0.99   & 2.4  & 3.6 & 4.8 \\[0,5 ex]
$\Gamma (D^0\to \pi^-) \; \left[10^{9 }s^{-1}\right]$	& $9.0 \pm 1.5$ & 2.06  & 1.99   & 4.8  & 7.1 & 9.6 \\[0,5 ex]
$\Gamma (D^+\to \rho^0) \; \left[10^{9 }s^{-1}\right]$	& $2.1 \pm 0.8$ & 2.25  & 3.36   & 1.2  & 3.4 & 2.1 \\[0,5 ex]
$\Gamma_L/\Gamma_T$					& --- 		& 1.30	& 1.55   & 0.67 & 0.91& 1.16 \\[0,5 ex]
$\Gamma_+/\Gamma_-$					& --- 		& 0.15	& 0.26   & ---  & --- & ---\\
&&&&&&\\						  
$A_1(0)$						& --- &  0.58	& 0.72	& ---  & 0.78	& 0.6 \\[0,5 ex]
$r_V$							& --- &  1.68 	& 1.26 	& ---  & 1.58	& 1.48 \\[0,5 ex]
$r_2$							& --- &  0.77	& 0.59  & ---  & 1.18	& 0.82\\
&&&&&&\\
\end{tabular}
\vspace*{0.5ex}
\caption{$D\to \pi/\rho\ell\bar\nu$ decay observables and form factor
parameters. We use $|V_{cd}| = 0.222$ \protect\cite{PDG2k}. \label{Dpi}}
\end{table}
\begin{table}
\begin{tabular}{lcddddd}
&&&&&&\\
\bf Parameter			& \bf Exp. \cite{PDG2k}	&\bf Model $\cal A$	& \bf Model $\cal B$	& \bf ISGW2 \cite{ISGW2}& \bf WSB \cite{WSB} &\bf MS\cite{MS}\\
&&&&&&\\
\hline
&&&&&&\\
$\Gamma (D\to \eta) \; \left[10^{9 }s^{-1}\right]$	& --- & 0.79 	& 0.95   & 1.5 $^{ a)}$ & --- & --- \\[0,5 ex]
$\Gamma (D\to \eta') \; \left[10^{9 }s^{-1}\right]$	& --- & 0.14	& 0.16   & 0.3 $^{ a)}$ & --- & --- \\[0,5 ex]
$\Gamma (D\to \omega) \; \left[10^{9 }s^{-1}\right]$	& --- & 2.23	& 3.35   & 1.2 & --- & --- \\[0,5 ex]
$\Gamma_L/\Gamma_T$					& --- & 1.31	& 1.55 	 & 0.68& --- & --- \\[0,5 ex]
$\Gamma_+/\Gamma_-$					& --- & 0.15	& 0.26 	 & --- & --- & --- \\
&&&&&&\\						  
$A_1(0)$						& --- & 0.41 	& 0.51 	& --- &	--- & --- \\[0,5 ex]
$r_V$							& --- & 1.68  	& 1.27    	& --- &	--- & --- \\[0,5 ex]
$r_2$							& --- & 0.77 	& 0.59      & --- &	--- & --- \\
&&&&&&\\
\end{tabular}
\vspace*{0.5ex}
\caption{$D\to \eta/\eta'/\omega\ell\bar\nu$ decay observables and form factor
parameters. We use $|V_{cd}| =0.222$ \protect\cite{PDG2k}. \\ $^{a)}$ A $\eta-\eta'$ mixing angle of $-20^\circ$ is
assumed. An angle of $-10^\circ$ would lead to 1.1 and 0.4.\label{Deta}}
\end{table}
\begin{table}
\begin{tabular}{lcddddd}
&&&&&&\\
\bf Parameter			& \bf Exp. \cite{PDG2k}	&\bf Model $\cal A$	& \bf Model $\cal B$	& \bf ISGW2 \cite{ISGW2}& \bf WSB \cite{WSB} &\bf MS\cite{MS}\\
&&&&&&\\
\hline
&&&&&&\\
$\Gamma (D_s\to K  ) \; \left[10^{9 }s^{-1}\right]$	& --- & 3.42	& 3.15   & 4.4 & --- & 6.4 \\[0,5 ex]
$\Gamma (D_s\to K^* ) \; \left[10^{9 }s^{-1}\right]$	& --- & 2.71	& 4.54   & 2.2 & --- & 3.9 \\[0,5 ex]
$\Gamma_L/\Gamma_T$					& --- & 1.24	& 1.51   & 0.76& --- & 1.21\\[0,5 ex]
$\Gamma_+/\Gamma_-$					& --- & 0.14	& 0.29   & --- & --- & --- \\
&&&&&&\\						      
$A_1(0)$						& --- & 0.43 	& 0.58	& --- &	--- & 0.57 \\[0,5 ex]
$r_V$							& --- & 1.86  	& 1.26  	& --- &	--- & 1.83 \\[0,5 ex]
$r_2$							& --- & 0.76 	& 0.56    & --- &	--- & 0.74\\
&&&&&&\\
\end{tabular}
\vspace*{0.5ex}
\caption{$D_s\to K/K^*\ell\bar\nu$ decay observables and form factor
parameters. We use $|V_{cd}| = 0.222$ \protect\cite{PDG2k}. \label{DsK}}
\end{table}
\endtable

\begin{table}
\renewcommand{\arraystretch}{1}
\begin{tabular}{lcddd}
\\
\bf Decay ratio      & \bf Exp. \cite{PDG2k}& \bf Model $\cal A$ & \bf Model $\cal B$  &\bf NRSX \cite{Neubert}\\
\\
\hline
\\
$\frac{\Gamma(B^0\to D^-\pi^+)}{\Gamma(B^0\to D^{*-}\pi^+)}$     & $1.09 \pm 0.17$  & 1.00 & 1.05  & 1.04 \\
\\
$\frac{\Gamma(B^0\to D^-\rho^+)}{\Gamma(B^0\to D^{*-}\rho^+)}$    & $1.16 \pm 0.62$  & 0.87 & 0.92 & 0.88\\
\\
\begin{tabular}{l}$\frac{\Gamma(B^0\to D^-D_s^+)}{\Gamma(B^0\to D^{*-}D_s^+)}$\\[2 ex] $\frac{\Gamma(B^+\to \overline{D}^0D_s^+)}{\Gamma(B^+\to \overline{D}^{*0}D_s^+)}$ \end{tabular}   & \begin{tabular}{c}$0.83 \pm 0.43$\\[2 ex]$1.08 \pm 0.56$\end{tabular} & 1.42  & 1.28 & 1.47 \\
\\
\begin{tabular}{l}$\frac{\Gamma(B^0\to D^-D_s^{*+})}{\Gamma(B^0\to D^{*-} D_s^{*+})}$\\[2 ex]$\frac{\Gamma(B^+\to \overline{D}^0D_s^{*+})}{\Gamma(B^+\to \overline{D}^{*0} D_s^{*+})}$ \end{tabular} & \begin{tabular}{c}$0.50 \pm 0.31$\\[2 ex]$0.33 \pm 0.19$\end{tabular} & 0.41  & 0.36 & 0.39 \\
\\
\end{tabular}
\vspace*{0.5ex}
\caption{Ratios of non-leptonic B meson decay rates}\label{nonlept_ratios}
\widetext
\end{table}

\begin{table}
\begin{tabular}{lcddd}
\\
\bf Decay mode      & \bf Exp. \cite{PDG2k} & \bf Model $\cal A$ & \bf Model $\cal B$  & \bf NRSX\cite{Neubert}\\
\\
\hline
\\
$B^0\to D^-\pi^+$     & $1.94 \pm 0.26$  & 2.21 & 1.97 & 1.94 \\[0,5 ex]
$B^0\to D^-\rho^+$    & $5.10 \pm 0.90 $ & 5.68 & 6.13 & 4.84 \\[0,5 ex]
$B^0\to D^{*-}\pi^+$  & $1.78 \pm 0.14 $ & 2.22 & 1.88 & 1.87 \\[0,5 ex]
$B^0\to D^{*-}\rho^+$ & $4.4 \pm 2.2$  	 & 6.57 & 6.67 & 5.48 \\
\\
$B^0\to D^-K^+$	      & --- & 0.17 & 0.15  & 0.15 \\[0,5 ex]
$B^0\to D^-K^{*+}$    & --- & 0.31 & 0.30  & 0.26 \\[0,5 ex]
$B^0\to D^{*-}K^+$    & --- & 0.16 & 0.14  & 0.14 \\[0,5 ex]
$B^0\to D^{*-}K^{*+}$ & --- & 0.36 & 0.34  & 0.32 \\
\\					  
$B^0\to D^-D^+$	      & --- & 0.53 & 0.49  & 0.23  \\[0,5 ex]
$B^0\to D^-D^{*+}$    & --- & 0.37 & 0.34  & 0.23  \\[0,5 ex]
$B^0\to D^{*-}D^+$    & --- & 0.39 & 0.41  & 0.17  \\[0,5 ex]
$B^0\to D^{*-}D^{*+}$ & $0.40 {+0.26 \atop -0.20}$ & 0.86 & 0.89 & 0.54  \\
\\
$B^+\to \overline{D}^0D_s^+$	& $7.9 \pm 2.4$  & 8.73  & 8.27 & 6.61 \\[0,5 ex]
$B^0\to D^-D_s^+$       	& $5.2 \pm 1.9$  & 8.73  & 8.27 & 6.61 \\[0,5 ex]
$B^+\to \overline{D}^0D_s^{*+}$ & $5.4 \pm 2.4 $ & 6.05  & 5.49 & 6.15  \\[0,5 ex]
$B^0\to D^-D_s^{*+}$    	& $6.5 \pm 3.2 $ & 6.05  & 5.49 & 6.15  \\[0,5 ex]
$B^+\to \overline{D}^{*0}D_s^+$ & $7.3 \pm 3.0 $ & 6.14  & 6.47 & 4.49 \\[0,5 ex]
$B^0\to D^{*-}D_s^+$    	& $6.2 \pm 2.2 $ & 6.14  & 6.47 &  4.49 \\	[0,5 ex] 
$B^+\to \overline{D}^{*0} D_s^{*+}$& $16.3 \pm 6.0$  & 14.68 & 15.24 & 15.8 \\[0,5 ex]
$B^0\to D^{*-} D_s^{*+}$ 	& $12.9 \pm 4.5$ & 14.68     & 15.24 & 15.8 \\
\\
\end{tabular}
\vspace*{0.5ex}
\caption{Non-leptonic B decay rates $\Gamma  \;[ns^{-1}]$. We use $|V_{cs}| = 0.975$, $|V_{ud}| = 0.975$, $|V_{us}| = 0.223$ \protect\cite{PDG2k} and  $|V_{cb}| = 0.034 \;({\cal A})$ resp. $|V_{cb}| =
0.035 \;({\cal B})$, as described in section \ref{semilep}. The results of \protect\cite{Neubert} have been adapted to the decay constants used in our calculation.}\label{nonlept}
\end{table}

\begin{table}
\begin{tabular}{lcdd}
\\
\bf Meson      & \bf Exp. \cite{PDG2k}  & \bf Model $\cal A$ & \bf Model $\cal B$ \\
\\\hline\\
$\pi$      & $130.7 \pm  0.1 \pm  0.36$  & 212 & 219\\[0,5 ex]
$K$       &  $159.8 \pm  1.4 \pm  0.44$  & 248 & 238\\[0,5 ex]
$D$        & $300 {+180+80 \atop -150-40}$   & 293 & 263\\[0,5 ex]
$D_s$      & $280 \pm 19 \pm 28 \pm 34$     & 315 & 284\\
\\
$\rho$     & $216 \pm 5$ \cite{Neubert}  & 470 & 717 \\[0,5 ex]
$\omega$   & $195 \pm 3$ \cite{Neubert}  & 472 & 726\\[0,5 ex]
$\phi$     & $237 \pm 4$ \cite{Neubert}  & 475 & 685\\[0,5 ex]
$K^*$      & $214 \pm 7$ \cite{Neubert}  & 467 & 695\\[0,5 ex]
$D^*$      & ---  & 339 & 409\\[0,5 ex]
$D_s^*$    & ---  & 378 & 445\\
\end{tabular}
\vspace*{0.5ex}
\caption{Decay constants of pseudoscalar and pseudovector mesons in MeV}\label{weakdec}
\end{table}


\end{document}